\documentclass[twocolumn,showpacs,preprintnumbers,amsmath,amssymb]{revtex4}
\usepackage{axodraw}
\usepackage{pstricks}
\usepackage{color}
\usepackage{graphicx}% Include figure files
\usepackage{dcolumn}% Align table columns on decimal point
\usepackage{bm}% bold math
\newcommand{\eq}[1]{eq.~(\ref{#1})}
\newcommand{\fig}[1]{Fig.~\ref{#1}}

\begin{document}
\preprint{TTP10-43}

\title{Yukawa coupling and anomalous magnetic moment of the muon:\\
       an update for the LHC era}
% Force line breaks with \\
%
\author{Andreas Crivellin${}^{1,2}$, Jennifer Girrbach${}^{1}$, and Ulrich Nierste${}^{1}$}
\affiliation{${}^{1}$Institut f\"ur Theoretische Teilchenphysik\\
               Karlsruhe Institute of Technology, 
               Universit\"at Karlsruhe, \\ D-76128 Karlsruhe, Germany.\\
               ${}^{2}$Albert Einstein Center for Fundamental Physics, Institute for Theoretical Physics,\\
               University of Bern, CH-3012 Bern, Switzerland.}%
\date{October 2010}
\begin{abstract}
  We study the interplay between a soft muon Yukawa coupling generated
  radiatively with the trilinear $A$-terms of the minimal supersymmetric
  standard model (MSSM) and the anomalous magnetic moment of the muon.
  In the absence of a tree-level muon Yukawa coupling the lightest smuon
  mass is predicted to be in the range between $600~\rm{GeV}$ and
  $2200\,\rm{GeV}$ at $2\sigma$, if the bino mass $M_1$ is below
  $1\,\rm{TeV}$. Therefore, a detection of a smuon (in conjunction with
  a sub-TeV bino) at the LHC would directly imply a nonzero muon Yukawa
  coupling in the MSSM superpotential. Inclusion of slepton flavor
  mixing could in principle lower the mass of 
  one smuonlike slepton below $600~\rm{GeV}$.
  However, the experimental bounds on
    radiative lepton decays instead strengthen the lower mass bound,
    with larger effects for smaller $M_1$, We also extend the analysis
  to the electron case and find that a light selectron close to the
  current experimental search limit may prove the MSSM electron Yukawa
  coupling to be nonzero.
\end{abstract}
                              %Classification Scheme.
\maketitle

\section{Introduction}

In the standard model (SM) the measured fermion masses determine the
values of the Yukawa couplings. Beyond the standard model, however, the
Yukawa sector is \emph{terra incognita}: In the minimal supersymmetric
standard model (MSSM), which we consider in this paper, the relations
between the masses $m_f$ and the Yukawa couplings $y_f$ depend on the
ratio of the two vacuum expectation values $\tan\beta = v_u/v_d$ and
further receive important radiative contributions from the soft
supersymmetry-breaking sector.  In the decoupling limit $M_{\rm SUSY}\gg
M_{A^0},M_{H^+},v_u$ the effective loop-induced Yukawa coupling arise in
an intuitive way from one Higgs coupling to sfermions, which involve
either the trilinear terms $A^f_{ij}$ ($f=u,d,\ell$ and $i,j=1,2,3$
label the generation) or the Higgsino mass parameter $\mu$ accompanied
by $y_f$, by integrating out the heavy SUSY particles
\cite{Banks:1987iu,Hall:1993gn,Hempfling:1993kv,Blazek:1995nv,Hamzaoui:1998nu,Isidori:2001fv,bcrs,gjnt}.
However, if no hierarchy between the sparticle mass
scale $M_{\rm SUSY} $, the Higgs masses, and the vevs is present, one
has to calculate the self-energy diagrams using exact diagonalization
(which correctly accounts for all powers of $A$-terms
and $\mu Y^f$).  This then leads to a finite renormalization of the
Yukawa couplings and mixing matrices of quarks
\cite{cgnw,Hofer:2009xb,Crivellin:2008mq} and leptons
\cite{Girrbach:2009uy}.  With the appropriate all-order resummations the
result for the Yukawa coupling reads:
\begin{equation}
y_{f_i} % \;=\; \frac{m_{f_i}}{v_{u,d} \, (1 + \Delta_{f_i}) } 
\;=\;  \frac{m_{f_i}-\Sigma^{f\,LR}_{ii,\,A}}{v_{u,d} \, (1
+ \Sigma^{f\,LR}_{ii,\,\mu}/(y_{f_i} v_{u,d}) ) }.\label{mqdel}
\end{equation}
In Eq.~(\ref{mqdel}) we have decomposed the self-energy $\Sigma^{f\,LR}_{ii}$
as $\Sigma^{f\,LR}_{ii,\,A}+\Sigma^{f\,LR}_{ii,\,\mu}$
as in \cite{Crivellin:2008mq}.
$\Sigma^{f\,LR}_{ii,\,\mu}$ is proportional to $\mu\,y_{f_i}$ and
$\Sigma^{f\,LR}_{ii,\,A}$ is the remaining part of the self-energy, in
which the chirality-flip does not stem from $y_{f_i}$, but e.g. from
$A^{f}_{ii}$. Equation~(\ref{mqdel}) is only correct for negligible flavor-mixing. Furthermore, since the size of $\Sigma^{f\,LR}_{ii,\,A}$
can be of the order of the light fermion masses, we do not even know if
the light fermions possess a tree-level Yukawa coupling at all, because
it might be possible that their masses are entirely generated by the
radiative contribution
\cite{Weinberg:1972ws,Lahanas:1982et,Buchmuller:1982ye,Nanopoulos:1982zm,Masiero:1983ph,Banks:1987iu,Babu:1989fg,Babu:1998tm,Borzumati:1999sp,Ferrandis:2004ri,Crivellin:2008mq,Crivellin:2009pa,Crivellin:2010gw}.
For a vanishing hard Yukawa coupling $y_f$ in the superpotential one has
$\Sigma^{f\,LR}_{ii,\,\mu}=0$ at the one-loop level.  However, the same
trilinear term $A^f_{ii}$ needed to generate the soft contribution
stemming from $\Sigma^{f\,LR}_{ii,\,A}$ in \eq{mqdel} also enters the
anomalous magnetic moment of the corresponding fermion
\cite{Borzumati:1999sp}.  The anomalous magnetic moment of the muon,
$a_\mu = \tfrac{1}{2}(g-2)_\mu$, is of special importance since its
precisely measured value deviates from the SM prediction by more than
$3\sigma$
\cite{Bennett:2006fi,Jegerlehner:2009ry,Davier:2009zi,Teubner:2010ah,Passera:2010ev,Prades:2009qp,Davier:2010nc}.
An improved $(g-2)_\mu $ experiment could decrease the experimental error by a factor of 4 \cite{NewExp}.
The interplay between $a_\mu$ and a radiatively generated muon mass was
already studied in \cite{Borzumati:1999sp}: If $A^\ell_{22}$ is adjusted
to reproduce $m_\mu$, one tends to overshoot the desired new-physics
contribution to $a_\mu$.

In the next section we will update the 1999 analysis of
\cite{Borzumati:1999sp} using present-day inputs from experiment and
theory. We extend this study by including the effects of slepton
mixing, which can lead to additional contributions proportional to $y_\tau$.  
With sufficiently heavy bino and smuon masses one can generate $m_\mu$
from the nondecoupling soft loop contribution
$\Sigma^{f\,LR}_{22,\,A}$ and simultaneously satisfy the $a_\mu$
constraint which disappears in the decoupling limit. Therefore, the
discoveries of these particles at the LHC will eventually permit to
rule out an entirely soft muon mass and instead establish a nonzero
$y_\mu$ in the superpotential. The aim of this paper is to quantify
this statement.  In addition, we investigate the electron case.

\boldmath
\section{Correlation between ${a_\mu}$ and  ${y_\mu}$}
\unboldmath
The magnetic dipole moment interaction relevant for $a_\mu$ is
given by 
\begin{equation}
 \frac{ie}{2m_\mu}F(q^2)\overline{u}(p_f)\sigma_{\mu\nu}q^\mu\epsilon^\nu u(p_i)
\end{equation}
where $q = p_f -p_i$ is the momentum and $\epsilon$ is the polarization vector of the
external photon. The anomalous magnetic dipole moment of the muon is then given as
\begin{equation}
 a_\mu =  F(q^2=0) .
\end{equation}
Experimentally, $a_\mu$ differs from its SM prediction by
$3.6\sigma$ \cite{Davier:2010nc}:
\begin{equation}
a^{\rm{exp}}_{\mu}-a^{\rm{SM}}_{\mu}=(28.7\pm8.0)\times10^{-10}.
\label{a-mu}
\end{equation}
In unbroken supersymmetric theories the gryomagnetic ratio for all
fermions is exactly $2$ \cite{Ferrara:1974wb}. Therefore, the
anomalous magnetic moment of the muon directly probes SUSY breaking. A
pleasant feature of supersymmetry, which distinguishes it from many
alternative theoretical frameworks, is that it can naturally explain
the observed deviation from the SM value
\cite{Grifols:1982vx,Ellis:1982by,Barbieri:1982aj,Kosower:1983yw,Yuan:1984ww,Carena:1996qa,Dedes:2001fv,Stockinger:2006zn,Marchetti:2008hw}. The usual
approach is to choose a suitable (large) value of the term $|y_{\mu}|
\mu v_u\approx m_\mu\mu\tan\beta$ \footnote{Definitions and sign
  conventions of SUSY parameters are as in \cite{Rosiek:1995kg}.}. In
order to achieve the right value for the anomalous magnetic moment,
the higgsino mass parameter $\mu$ must be positive and large values
for $\tan\beta\gtrsim 10$, the ratio of the two vacuum expectation
values, are favored. While large-$\tan\beta$ scenarios are also
motivated by the GUT relation $|y_t|=|y_b|$, problems in processes
like $b\to s\gamma$, $B_{d,s}\to \mu\bar\mu$ and $B\to (D)\tau \nu$
can occur, due to the parametric enhancement by $\tan\beta$
\cite{bcrs,ip,ntw,gjnt}. 
In mSUGRA and the constrained MSSM, $B_{d,s}\to \mu\bar\mu$ and  the
anomalous magnetic moment of the muon are correlated, limiting the
possible size of $a^{\rm{SUSY}}_{\mu}$\cite{Dedes:2001fv}.  Therefore,
if $\tan\beta$ is large, the nonstandard Higgses have to be heavy in
the CMSSM \cite{Eriksson:2008cx}.

However, there exists also a second, less studied way in the MSSM to
account for the anomalous magnetic moment of the muon: The entry in the
slepton mass matrix involving the trilinear
SUSY-breaking terms, $v_d A^\ell_{22}+v_u
A^{\ell\prime}_{22}$, can also reproduce the desired effect without
influencing quark decays or the Higgs potential. This possibility is
realized in models with radiative generation of fermion masses
\cite{Buchmuller:1982ye,Borzumati:1999sp,Ferrandis:2004ri,Crivellin:2008mq,Crivellin:2009pa}.
In these models the trilinear terms are chosen in such a way that they
generate the light fermion masses of the first and second generation,
while the corresponding tree-level Yukawa couplings are zero. The
radiative generation of fermion masses has several advantages compared
to the general (and also the minimally flavor-violating MSSM)
\cite{Borzumati:1999sp,Crivellin:2008mq, Crivellin:2009pa}:
\begin{figure}[t]
\includegraphics[width=0.7\linewidth]{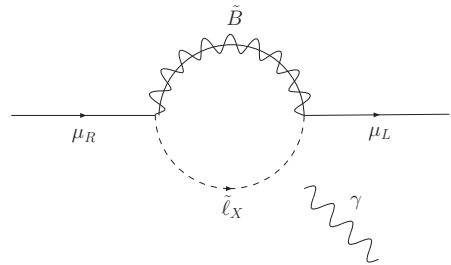}
\caption{Self-energy contribution $\Sigma^{\ell\,LR}_{22,\,A}$
  constituting $m_\mu$ for $y_\mu=0$.  $\tilde \ell_X$, $X=1,\ldots 6$
  are the charged-slepton mass eigenstates.  Diagram
    with the attached photon contributes to $a_\mu$.\label{fig}}
\end{figure}
\begin{itemize}
  \item The otherwise approximate $[U(2)]^5$ flavor symmetry in the
    Yukawa sector becomes exact.
  \item It is minimally flavor-violating when only the first two
    generations are involved due to a natural alignment between the
    $A$-terms and the effective Yukawa couplings. In contrast to the
    naive definition of MFV (switching FCNCs mediated by gluinos and
    neutralinos off) the Yukawa sector is renormalisation-group invariant
    thanks to Symanzik's nonrenormalization theorem.
  \item There is no SUSY CP problem because the phases of the
    $A$-terms and the Yukawa couplings of the first two generations
    are automatically aligned.  In addition, the phase of $\mu$
    essentially only enters at the two-loop level (apart from a very
    small neutralino mixing effect) \cite{Borzumati:1999sp}.
\end{itemize}
The shift in the anomalous magnetic moment depends only on the slepton
and bino masses and is positive \cite{Borzumati:1999sp}. Note that
this is an advantage of radiative mass generation
compared to  models with very large $\tan\beta$
\cite{Dobrescu:2010mk,Altmannshofer:2010zt}, in which 
the discrepancy between $a^{\rm{exp}}_{\mu}$ and the theory prediction
becomes larger. In the following we set $y_\mu=0$ and examine the 
phenomenological consequences for the sparticle spectrum.

As already examined in \cite{Borzumati:1999sp} (based on the
analysis in \cite{Gunion:1987qv}) vacuum stability (VS) is
critical for the muon. Since the constraints from VS are
nondecoupling they are equally valid for any value of $M_{\rm SUSY} $. 
If the muon mass is generated radiatively, the vacuum cannot be
absolutely stable; only meta-stability is possible.  Using the analytic
trilinear term $A_{22}^\ell$ VS can only be satisfied for very small
values of $\tan\beta\approx 1$. However, such a low value of $\tan\beta $ 
causes problems with the
pertubativity of the top Yukawa coupling and the mass of the lightest 
Higgs boson.
This issue can be avoided by using the nonanalytic $A$-term
\cite{Haber:2007dj,Rosiek:1995kg,Demir:2005ti} $A^{\ell\prime}_{22}$ which
comes with $v_u$ in the slepton mass matrix
\cite{Demir:2005ti}. Note that only the
combination $A_{22}^\ell v_d$ or $A^{\ell\prime}_{22} v_u$ enters in the
off-diagonal element of the smuon mass matrix. One can only
distinguish the different types of $A$-terms by considering Higgs-mediated
processes.

\begin{figure*}
\includegraphics[width=0.43\textwidth]{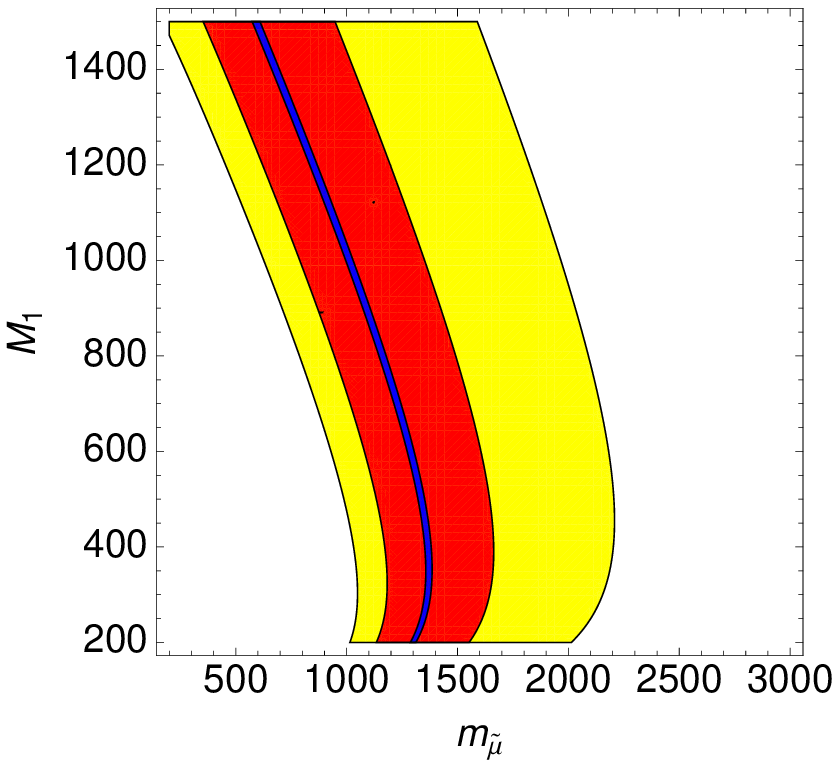}
\includegraphics[width=0.43\textwidth]{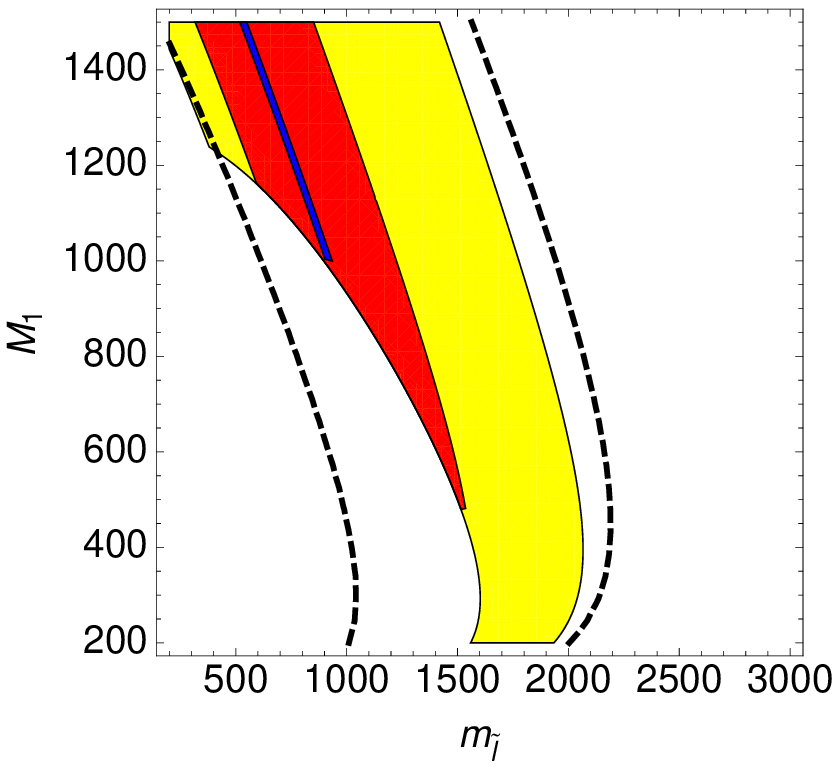}

\includegraphics[width=0.43\textwidth]{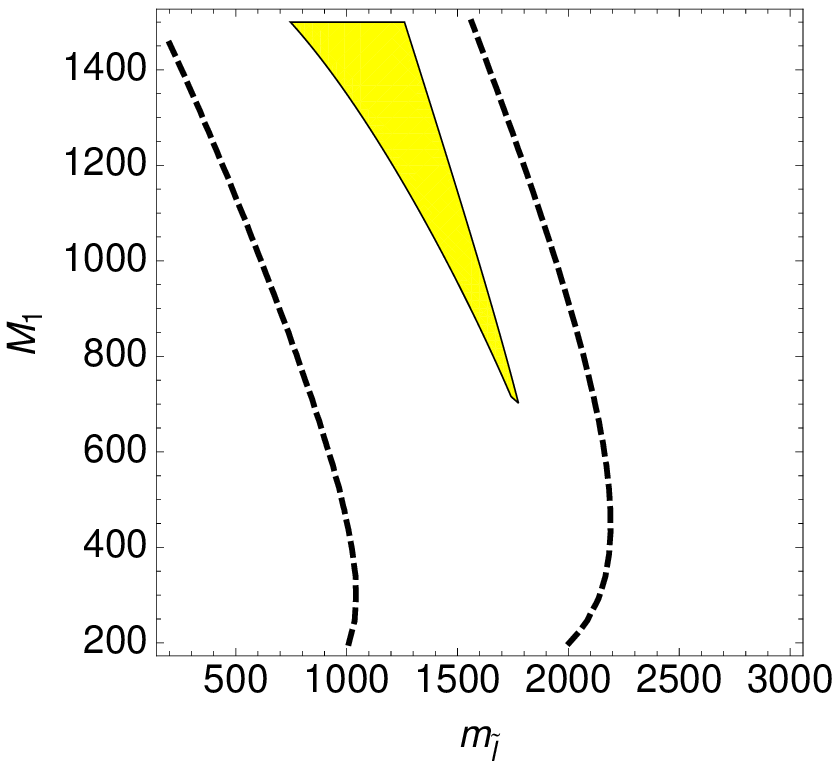}
 \includegraphics[width=0.4\textwidth]{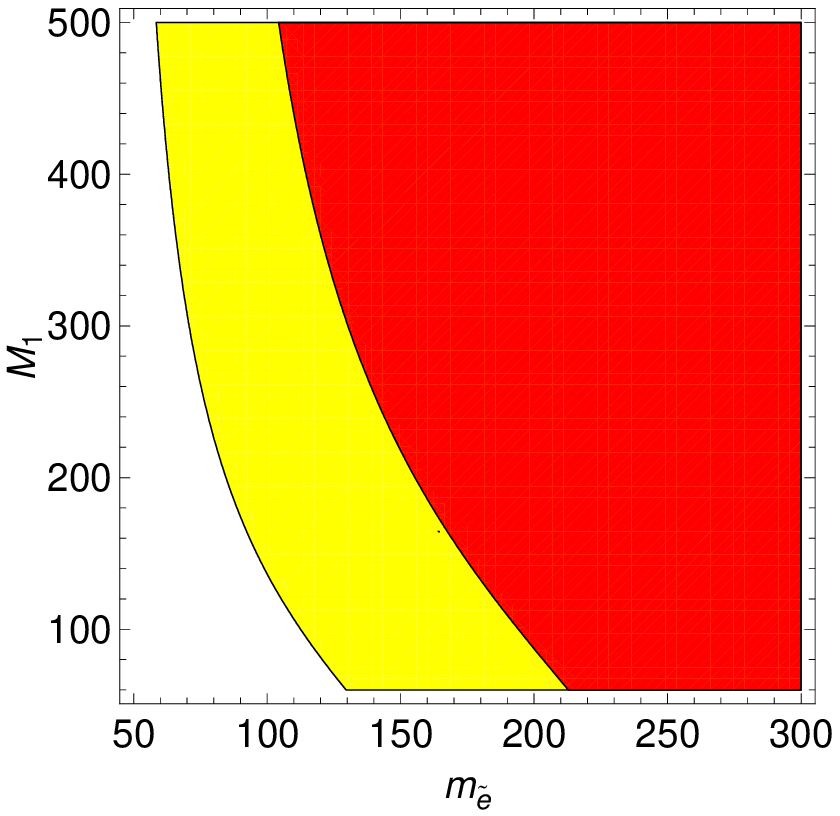}
\caption{Top left: Allowed region in the $M_1$-$m_{\tilde \mu}$  plane
  assuming that the muon Yukawa coupling is generated radiatively by
  $v_d A^\ell_{22}$ and/or $v_u A^{\ell\prime}_{22}$.  Here $m_{\tilde \mu}$
  is the lighter smuon mass. Yellow (lightest): $a_{\mu}\pm2\sigma$,
  red: $a_{\mu}\pm1\sigma$, blue (darkest): $a_{\mu}$. Top right: Allowed
  region in the $M_1$-$m_{\tilde \ell}$ plane including lepton flavor violation
  with $\delta_{23}^{\ell\,RR}=0.3$ and the constraint from $\tau\to\mu\gamma$ (black dashed for $\delta_{23}^{\ell\,RR}=0$). Down left: same with $\delta_{23}^{\ell\,RR}=0.5$. Down right: Allowed
  region in the $M_1$-$m_{\tilde e}$ plane assuming that the electron
  Yukawa coupling is generated radiatively with $v_d A^\ell_{11}$ and/or
  $v_u A^{\ell\prime}_{11}$. Yellow (lightest): $a_e\pm2\sigma$, red:
  $a_e\pm1\sigma$.}
    \label{fig:gm2region} 
\end{figure*}

In the presence of flavor-violating elements in the slepton mass matrix
it is possible to generate part of the muon (and electron) mass radiatively
via couplings involving $y_\tau$. The slepton mass eigenstates in the
diagram of Fig.~\ref{fig} are then linear combinations of
$\tilde\mu_L$, $\tilde\mu_R$, $\tilde\tau_L$, and $\tilde\tau_R$ (and
possibly also of $\tilde e_L$, $\tilde e_R$).  However, by attaching a
photon to the charged-slepton line one obtains the corresponding
contribution to $a_\mu$ \footnote{flavor-mixing in magnetic and
electric dipole moments in the MSSM is studied in the
literature, see e.g.~\cite{Masina:2002mv,Hisano:2008hn}.}.
In addition, no chargino diagram contributes due to
the absence of a tree-level Yukawa coupling. Neglecting mixing between the bino
and neutral wino the
magnetic moment is given as
\begin{widetext}
\begin{align}
  a_\mu = & m_\mu \frac{\alpha_1}{2 \pi} M_1 \sum_{X =1}^6
\Re\left(Z_L^{2X}Z_L^{5X\ast}\right)m_{\tilde{\ell}_X}^2 D_0(M_1^2,m_{\tilde{\ell}_X}^2,m_{\tilde{\ell}_X}^2,m_{
\tilde{\ell}_X}^2),\label{equ:amuBino}
\end{align}
\end{widetext}
where $D_0$ is a loop function as defined in the appendix of
\cite{Crivellin:2008mq}. 
Further the following condition must be fulfilled:
\begin{align}
 m_\mu\stackrel{!}{=}
\frac{\alpha_1}{4\pi}M_1\sum_{X=1}^{6}Z_L^{2X}Z_L^{5X*}B_0(M_1^2,m_{\tilde{
\ell}_X} ^{2}).\label{equ:mMuonradiativ}
\end{align}
We choose the diagonal elements of the slepton mass matrix to be equal
such that Eq.~(\ref{equ:mMuonradiativ}) implicitly determines the off-diagonal
elements.  Assuming that the discrepancy in Eq.~(\ref{a-mu}) can be explained
within supersymmetry we can determine the allowed region in parameter
space.  The result for the case without flavor mixing
is shown in the top left plot of
Fig.~\ref{fig:gm2region}: We see that a model with radiative
generation of the muon mass predicts a lightest smuon with mass
approximately between $600~\rm{GeV}$ and $2200~\rm{GeV}$ unless $M_1$ is
heavier than $1~\rm{TeV}$.

With the inclusion of lepton-flavor violation it is in principle
possible to weaken this bound, because e.g.\
  $\widetilde{\mu}$--$\widetilde{\tau}$ mixing lowers the mass of one
  smuonlike mass eigenstate. However the effect is limited in size,
  because the correlation between a radiative muon mass and $a_\mu$
  stemming from the diagrams in \fig{fig} stays intact.  Moreover, the
flavor-changing elements of the slepton mass matrix are tightly
constrained by the radiative lepton decays $\ell_j\to\ell_i\gamma$.
For a recent analysis of the bounds on the dimensionless
  quantities $\delta^{\ell\,XY}_{ij}$, $X,Y=L,R$, parametrizing the
  off-diagonal elements of the slepton mass matrix we refer to
  \cite{Girrbach:2009uy}.  The constraints from $\ell_j\to\ell_i\gamma$
are weakest for $\delta^{\ell\,RR}_{23}$ which we have
kept nonzero in our analysis. The
  remaining elements are tightly constrained and negligible for our
  purpose.  In the following we focus on the lighter slepton mass
  eigenstate. We have checked that the smuon component of this
  eigenstate is indeed larger than the stau component, although almost
  an equal mixture of $\widetilde{\mu}_R$ and $\widetilde{\tau}_R$ is
  possible. That is, this slepton tends to decay into muons and would be
  identified as a ``smuon'' rather than a ``stau'' at the LHC. We
  observe that the allowed area in Fig.~\ref{fig:gm2region}
actually shrinks when
  $\widetilde{\mu}$--$\widetilde{\tau}$ mixing is included.  The reason
  for this is that the combined effect of $\delta^{\ell\,RR}_{23}$ and
  $\delta^{\ell\,LR}_{22}$, which is large to account for the radiative
  muon mass, mimics an effective element $\delta^{\ell\,LR}_{23}$
 which is severely constrained from $\tau \to
  \mu\gamma$.  In the second and third plot in
  Fig.~\ref{fig:gm2region} we choose exemplarily
  $\delta_{23}^{\ell\,RR}=0.3$ and $\delta_{23}^{\ell\,RR}=0.5$ and
 include the constraint $BR(\tau\to\mu\gamma)<4.4\times
  10^{-8}$ \cite{Benitez:2010gm} together with the
    condition for a radiative muon mass in Eq.~(\ref{equ:mMuonradiativ}).  We
  recognize the two abovementioned effects: with a
  large $\tilde\mu_L-\tilde\mu_R-\tilde\tau_R$ mixing the allowed area
  moves to the left because of a lighter
  smuom mass eigenstate.  However,
  $\tau\to\mu\gamma$ forbids nonvanishing $\delta^{\ell\,RR}_{23}$
  combined with a large $\delta^{\ell\,LR}_{22}$ for too light smuon
  masses and cuts the region corresponding to lighter
    sparticles out.  Thus, the interesting lower
limit on the smuon mass in a world with soft muon
  Yukawa coupling stays intact in the case of flavor mixing.  This is a
very clean and strong prediction which gains special importance in the
light of forthcoming LHC results: Since the LHC is only sensitive to
light sleptons with masses $m_{\tilde \ell} \leq 300\, \rm{GeV}$ at
$30~\rm{fb}^{-1}$\cite{Andreev:2004qq,delAguila:1990yw,Baer:1995va} a
detection of a smuon (in conjunction with a bino discovery or some upper
bound on $M_1$) would directly disprove the hypothesis of a radiatively
generated muon mass. Stated positively, a sufficiently light smuon will
imply a nonzero Yukawa coupling in the MSSM superpotential. Beyond the
MSSM, there is also the possibility of additional radiative
contributions from sparticles with very high masses, e.g.\ from the
messenger sector of gauge-mediated SUSY-breaking (see
\cite{Dobrescu:2010mk} and references therein). In such wider scenarios
the question of zero or nonzero $y_\mu$ might profit from additional
information gained from $B_{d,s}\to \mu^+\mu^-$ and $B^+\to
\mu^+\nu_\mu$ measured at the LHC and a super-B factory, respectively.

It is often stated that $a_\mu$ favors positive values of $\mu$ which
is especially true in the large-$\tan\beta$ case. However, the
inclusion of the trilinear $A$-terms can compensate the effect of the
$\mu$-term in the off-diagonal elements of the smuon mass matrix. This
permits the possibility of negative values of $\mu$ which would
otherwise be ruled out by the anomalous magnetic moment of the muon.

The discussion above applies as well to the electron and its Yukawa
coupling. However, even though the anomalous magnetic moment of the
electron is measured very precisely \cite{Aoyama:2007mn}, it is used
to determine $\alpha$. Therefore, in order to use the anomalous
magnetic moment of the electron to put bounds on new physics
parameters we need an independent determination of $\alpha$
\cite{Girrbach:2009uy}. The second best way to measure the fine
structure constant is from a Rubidium atom experiment
\cite{Cadoret:2008st}. Using this information we can qualitatively
make the same statements as in the muon case. However, quantitatively
the constraints are weaker due to the smallness of the electron mass
and the uncertainty coming from the second-best measurement of
$\alpha$ (see Fig.~\ref{fig:gm2region} down right).
 
\section{Conclusions}

The anomalous magnetic moment of the muon can be seen as a probe of a
tree-level muon Yukawa coupling.  In this paper we have performed an
updated analysis of the consequences of a radiatively generated muon
mass. We have found that this model is very predictive
if we use the new results for the anomalous magnetic moment of the muon:
The smuon mass must lie between $600~\rm{GeV}$ and $2200~\rm{GeV}$ for
$M_1<1\rm{TeV}$.  The inclusion of lepton-flavor violation does not
significantly change the picture.  This is mainly due to the fact that
the same diagrams also occur for the magnetic moment. In our analysis we
have included a nonvanishing $\delta_{23}^{\ell\,RR}$,
which cannot be constrained from $\tau\to\mu\gamma$ alone. In principle,
this could have decreased the lower bound
on the smuon mass, but in conjunction with a large
$\delta_{22}^{\ell\,LR}$ from a radiative muon mass, the constraint from
$\tau\to\mu\gamma$ forbids this possibility.
Consequently, the inclusion of lepton flavor violation cannot lower the
bound of approximately 600~GeV for the mass of the
  lightest smuonlike slepton.  If this smuon is found to be
lighter, the observed muon mass cannot entirely stem from the soft
SUSY-breaking sector.  Consequently, within the MSSM we
must then have a nonzero Yukawa coupling $y_\mu$ in the superpotential.
Therefore, we conclude that the high-$p_T$ experiments at the LHC can
shed light on the question whether $y_\mu$ is zero or not if they
discover a smuon.

%%%%%%%%%%%%%%%%%%%%%%%%%%%%%%
{\it Acknowledgments.}--- This work is supported by BMBF Grant No.05H09VKF
and by the EU Contract No.~MRTN-CT-2006-035482, \lq\lq
FLAVIAnet''. Andreas Crivellin and Jennifer Girrbach acknowledge the
financial support by the State of Baden-W\"urttemberg through
\emph{Strukturiertes Promotionskolleg Elementarteilchenphysik und
  Astroteilchenphysik}\ and the \emph{Studienstiftung des deutschen
  Volkes}, respectively. A.C.~is partially supported by the Swiss
National Foundation.  The Albert Einstein Center for Fundamental Physics
is supported by the ``Innovations- und Kooperationsprojekt C-13 of the
Schweizerische Universit\"atskonferenz SUK/CRUS.''

\bibliography{muon}% Produces the bibliography via BibTeX.

\end{document}